\documentclass[print]{mn2e}
\usepackage{epsfig}

\title[Effect of {\it r-}mode instability]{The effect of {\it r-}mode instability on the evolution of isolated strange stars}

\author[Xiao-Ping Zheng, Yun-Wei Yu and Jia-Rong Li]{Xiao-Ping Zheng,\thanks{zhxp@phy.ccnu.edu.cn (XPZ)} Yun-Wei Yu\thanks{yuyw@phy.ccnu.edu.cn (YWY)} and Jia-Rong Li
\\ Institute of Astrophysics, Huazhong Normal
University, Wuhan 430079, China}

\date{Accepted 2006 March 10. Received 2006 March 3; in original form 2006 January 8}
\pagerange{\pageref{firstpage}--\pageref{lastpage}} \pubyear{2006}

\begin{document}
\label{firstpage} \maketitle

\begin{abstract}

We studied the evolution of isolated strange stars (SSs)
synthetically, considering the influence of {\it r-}mode
instability. Our results show that the cooling of SSs with
non-ultrastrong magnetic fields is delayed by heating due to {\it
r-}mode damping for millions of years, while the spin-down of the
stars is dominated by gravitational radiation (GR). Especially for
the SSs in a possible existing colour-flavour locked (CFL) phase,
the effect of {\it r-}mode instability on the evolution of stars
becomes extremely important because the viscosity, neutrino
emissivity and specific heat involving pairing quarks are blocked.
It leads to the cooling of these colour superconducting stars being
very slow and the stars can remain at high temperature for millions
of years, which differs completely from previous understanding. In
this case, an SS in CFL phase can be located at the bottom of its
{\it r-}mode instability window for a long time, but does not
spin-down to a very low frequency for hours.
\end{abstract}


\section{INTRODUCTION}

The {\it r-}modes in a perfect fluid star succumb to gravitational
radiation (GR) driven Chandrasekhar-Friedman-Schutz instability for
all rates of stellar rotation, and they arise due to the action of
the Coriolis force with positive feedback increasing GR (Andersson
1998; Friedman \& Morsink 1998). However, actually, the viscosity of
stellar matter can hold back the growth of the modes effectively.
Based on the competition between the destabilizing effect of GR and
the damping effect of viscosity, Owen et al. (1998) modelled the
evolution of the {\it r-}modes and described the spin-down of
neutron stars as losing angular momentum to GR. Afterwards, Ho \&
Lai (2000) improved the model considering the conservation of wave
action and containing the magnetic braking due to magnetic dipole
radiation (MDR).
\par
The viscosity of stellar matter is decisive for the {\it r-}mode
evolution and hence for the spin-down and cooling of the star. The
bulk viscosity of strange quark matter in the normal phase is
several orders of magnitude stronger than that of normal nuclear
matter. As a consequence, the instability in a strange star (SS) can
only be driven at relatively lower temperature ($<10^{8}$K), which
means not very early ages, differing from the case of a neutron star
in which the instability acts immediately at the birth of the star
(Owen et al. 1998; Watts \& Andersson 2002). Thus Madsen (2000)
argued the evolution of SSs can be very different from that of
neutron stars. Andersson, Jones \& Kokkotas (2002) calculated the
evolution of SSs thoroughly using the model of Ho \& Lai (2000) (but
without a magnetic field) in the presence of accretion to point out
that SSs may be persistent sources of GR. However, most works always
focused on the stars with {\it r-}mode instability as sources of GR,
although the authors also claimed that the {\it r-}modes could
change the spin evolution of the stars due to GR and induce heating
due to viscous dissipation of the modes.
\par
We also note the work of Gusakov, Yakovlev, \& Gnedin (2005) who
studied the thermal evolution of a pulsating neutron star, taking
into account the pulsation damping due to viscosity and accompanying
heating of the star. This idea leads us to pay attention to the
effect of the {\it r-}mode instability on the evolution of compact
stars. As shown in fig. 5 in Watts \& Andersson (2002), by heating
due to {\it r-}mode damping, the cooling of a nascent neutron star
is delayed in the first few decades. In addition, the calculation in
Andersson et al. (2002) implies that the history of the {\it
r-}modes in isolated SSs could be longer than neutron stars.
According to the above works, we consider that a careful
investigation of the effect of the {\it r-}mode instability on the
evolution of SSs is an interesting topic, which we will focus on in
this paper. As expected, our results show that, although the {\it
r-}modes may be too small to induce a detectable GR at the age of a
million years, the accompanying heating effect still can not be
ignored in SSs.
\par
In particular, our attention is attracted by the same consideration
of colour superconducting SSs. For strange quark matter at a
sufficiently high density, phenomenological and microscopic studies
predict it could undergo a phase transition into a colour
superconducting state, such as the typical cases of two-flavour
colour superconductivity (2SC) and the colour-flavour locked (CFL)
phase (Shovkovy 2005; Alford 2004). Theoretical approaches also
concur that the pairing gap ($\Delta$) in the quark spectrum is
approximately $\sim100$MeV for baryon densities existing in the
interior of compact stars. In this case, the reaction rates
involving two quarks are reduced by a factor $\exp(-2\Delta/k_{\rm
B}T)$, assuming equal behaviour for all quark flavours. Because the
quarks in the 2SC phase pair in part, the corresponding SSs will
have similar behaviours to SSs in normal phase (NSS hereafter;
Blaschke, Kl$\rm\ddot{a}$hn, \& Voskresensky 2000; Madsen 2000; Yu
\& Zheng 2006). Therefore, here we only account for SSs in CFL phase
(CSS hereafter) with $\Delta=100$MeV, in which all flavour quarks
pair. As a consequence, all of the viscosity, the neutrino
emissivity and the quark specific heat of CSSs are reduced. Thus the
effect of the {\it r-}mode instability on the evolution of CSSs may
become more important. As shown in Section 5, the dramatic variation
in the stellar evolution changes our previous understanding of CSSs
that the stars are very cold and slow revolving objects (Blaschke et
al 2000; Madsen 2000).
\par
This paper is organized as follows: In Section 2 and 3 we recall the
{\it r-}mode instability and the model of spin evolution due to
gravitational and magnetic braking respectively. The thermal
evolution equation of SSs considering the heating due to the viscous
dissipation of the {\it r-}modes is given in Section 4. Section 5
presents the numerical results of the evolution of NSSs and CSSs.
Finally, we give the conclusion and our discussions in the last
section.

\section{The {\it r}-Mode Instability}

The {\it r-}modes of rotating Newtonian stars are generally defined
to be solutions of the perturbed fluid equations having (Eulerian)
velocity perturbations of the form (Lindblom, Owen \& Morsink 1998)
\begin{equation}
 \delta \vec{v} = \alpha R \Omega \left({r \over R}\right)^l \vec{Y}^{B}_{l\,m} e^{i\omega
t},
\end{equation}
where $R$ and $\Omega$ are the radius and angular velocity of the
unperturbed star, $\alpha$ is an arbitrary constant considered as
the amplitude of the {\it r-}modes (Owen et ale 1998), and
$\vec{Y}^{B}_{l\,m}$ is the magnetic-type vector spherical harmonic
defined by
\begin{equation}
 \vec{Y}^{B}_{l\,m}= [l(l+1)]^{-1/2}r\vec{\nabla} \times(r
\vec{\nabla}Y_{l\,m}).
\end{equation}
Papaloizou \& Pringle (1978) first showed that the Euler equation
for {\it r-}modes determines the frequencies as
\begin{equation}
 \omega = - {(l-1)(l+2)\over l+1}\Omega.
 \end{equation}
Further use of the Euler equation (as first noted by Provost,
Berthomieu \& Rocca 1981) in the barotropic case determines that
only $l=m$ {\it r-}modes exist and that $\delta \vec{v}$ must have
the radial dependence given by equation (1). These modes represent
large-scale oscillating currents that move (approximately) along the
equipotential surfaces of the rotating star. The density
perturbation associated with the {\it r-}modes can be deduced by
evaluating the inner product of $\vec{v}$ (the unperturbed fluid
velocity) with the perturbed Euler equation, and the equation for
the perturbed gravitational potential (Lindblom et ale 1998):
\begin{equation}
\begin{array}{cc}
 \delta \rho = \alpha R^2\Omega^2 \rho {d\rho\over dp}\qquad\qquad\qquad\qquad\\
\times\left[{2 l\over 2l+1}\sqrt{l\over l+1}\left({r\over
R}\right)^{l+1}+\delta\Psi(r)\right]Y_{l+1\,l} \,e^{i\omega t}.
\end{array}
\end{equation}
The quantity $\delta\Psi$ is proportional to the perturbed
gravitational potential $\delta\Phi$ and is the solution to the
ordinary differential equation
\begin{equation}
\begin{array}{cc}
 {d^2\delta \Psi(r)\over dr^2} + {2\over r} {d\delta
\Psi(r)\over dr} +\left[4\pi G\rho {d\rho\over dp}- {(l+1)(l+2)\over
r^2}\right]\delta \Psi(r) \nonumber\\
\qquad = -  {8\pi G l\over 2l+1} \sqrt{l\over l+1} \rho {d\rho\over
dp}\left({r\over R}\right)^{l+1}.
\end{array}
\end{equation}
\par
The {\it r-}modes evolve with time dependence $e^{i(\omega
+i/\tau)t}$ as a consequence of ordinary hydrodynamics and the
influence of the various dissipative processes. The real part of the
frequency of these modes, $\omega$, is given in equation (3), while
the imaginary part $1/\tau$ is determined by the effects of GR,
viscosity, etc.  The simplest way to evaluate $1/\tau$ is to compute
the time derivative of the energy $\tilde{E}$ of the mode (as
measured in the rotating frame). $\tilde{E}$ can be expressed as a
real quadratic functional of the fluid perturbations:
\begin{equation}
 \tilde{E}={1\over
2}\int\left[\rho\delta\vec{v}\cdot\delta\vec{v}^* +\left({\delta
p\over \rho}-\delta\Phi\right)\delta\rho^*\right]d^3x.
\end{equation}
Thus, the time derivative of $\tilde{E}$ is related to the imaginary
part of the frequency $1/\tau$, which can be conveniently decomposed
into two parts associated with GR and viscosity, respectively. It
reads
\begin{equation}
 {d\tilde{E}\over dt}= -{2\tilde{E}\over \tau}= -{2\tilde{E}\over \tau_{\rm g}} -{2\tilde{E}\over \tau_{\rm v}}.
 \end{equation}
For $n=1$ polytrope, the time-scale for GR
 is calculated by Lindblom, Mendell \& Owen (1999)
\begin{equation}
\tau_{\rm g}=-3.26(\Omega/\sqrt{\pi
G\bar{\rho}})^{-6}\hspace{0.1cm}\textrm{s},
\end{equation}
where $\bar{\rho}$ is the mean density of the star. The viscous
time-scale is contributed by shear and bulk viscosity as
\begin{equation}
\tau_{\rm v}=(\tau_{\rm sv}^{-1}+\tau_{\rm bv}^{-1})^{-1}.
\end{equation}
For strange quark matter, we can give (Lindblom et al. 1999; Madsen
2000)
\begin{equation}
\tau_{\rm sv}=5.41\times10^{9}\alpha_{\rm
c,0.1}^{5/3}T_{9}^{5/3}\hspace{0.1cm}\textrm{s}
\end{equation}
\begin{equation}
\tau_{\rm bv}=0.886(\Omega/\sqrt{\pi
G\bar{\rho}})^{-2}T_{9}^{-2}m_{100}^{-4}\hspace{0.1cm}\textrm{s},
\end{equation}
where $\alpha_{\rm c,0.1}$, $m_{100}$, and $T_{9}$ are the strong
coupling in units of 0.1, the mass of strange quark in units of
100MeV and the interior temperature of the stars in units of
$10^{9}$K, respectively. When these viscosities due to quark
reactions are blocked in the CFL phase, the dissipation of the {\it
r-}modes will be mainly dominated by a smaller shear viscosity due
to electron-electron scattering. The corresponding time-scale is
(Madsen 2000)
\begin{equation}
\tau_{\rm sv}^{\rm ee}=2.95\times 10^{7}(\mu_{\rm e}/\mu_{\rm
q})^{-14/3}T_{9}^{5/3}\hspace{0.1cm}\rm s.
\end{equation}
From equation (7), it can be seen that the {\it r-}modes are
unstable if $\tau_{\rm g}^{-1}+\tau_{\rm v}^{-1}<0$. In this case, a
small perturbation will lead to an unbounded growth of the modes.
The competition between the destabilizing effect of GR that is
dependent on spin frequency and the damping effect of the
temperature-dependent viscosity gives an instability window in the
$\nu-T$ plane, as shown by the shadows in the following figures. We
can see that, compared with the window of an NSS, the window of a
CSS expands significantly because its viscosity involving quarks are
blocked.

\section{spin evolution}
We employ a simple phenomenological spin evolution model proposed by
Ho \& Lai (2000), which is analogous to that devised by Owen et al.
(1998). In this case, the total angular momentum of a star can be
obtained as the sum of the bulk angular momentum and the canonical
angular momentum of the {\it r-}modes $J_{c}$
\begin{equation}
J=I\Omega+J_{c},
\end{equation}
where $I=\tilde{I}MR^{2}$ ($\tilde{I}=0.261$ for $n=1$ polytrope)
represents the moment of inertia of the star. The canonical angular
momentum of an {\it r-}mode can be expressed in terms of the
velocity perturbation $\delta\vec{v}$ by (Friedman \& Schutz 1978)
\begin{equation}
 J_c = -{l\over2 (\omega+l\Omega)}\int \rho\,
\delta\vec{v}\cdot\delta\vec{v}^{\,*} d^3x.
\end{equation}
For the (dominant) $l=m=2$ current multipole of the {\it r-}modes,
the above expression reduces (at the lowest order in $\Omega$) to
\begin{equation}
J_{c}=-\frac{3}{2}\alpha^{2}\tilde{J}MR^{2}\Omega,
\end{equation}
where  $\tilde{J}=1.635\times 10^{-2}$ for $n=1$ polytrope. The
canonical angular momentum of the mode can increase through GR back
reaction and decrease by transferring angular momentum to the star
through viscosity:
\begin{equation}
\frac{d{J}_{c}}{dt}=-\frac{2J_{c}}{\tau_{\rm
g}}-\frac{2J_{c}}{\tau_{\rm v}}.
\end{equation}
On the other hand, adding the reverse angular momentum transferred
from the {\it r-}mode, the bulk angular momentum of the star
decreases. At the same time, the star also undergoes braking by MDR.
Thus the rotation of the star is determined by
\begin{equation}
\frac{d(I\Omega)}{dt}=\frac{2J_{c}}{\tau_{\rm
v}}-\frac{I\Omega}{\tau_{\rm m}},
\end{equation}
where $\tau_{\rm m}=1.69\times 10^{9}B_{12}^{-2}(\Omega/\sqrt{\pi
G\bar{\rho}})^{-2} \rm s$ is the magnetic braking time-scale, and
$B_{12}$ is the magnetic field intensity in units of $10^{12}$G.
\par
Submitting equation (15) into equations (16) and (17), we can obtain
the coupled evolution equations of the amplitude $\alpha$ of the
{\it r-}mode and the angular velocity $\Omega$:
\begin{equation}
\frac{d\alpha}{dt}=-\alpha\left(\frac{1}{\tau_{\rm
g}}+\frac{1-\alpha^{2}Q}{\tau_{\rm v}}-\frac{1}{2\tau_{\rm m}}\right
),
\end{equation}
\begin{equation}
\frac{d\Omega}{dt}=-\Omega\left(\frac{2\alpha^{2}Q}{\tau_{\rm
v}}+\frac{1}{\tau_{\rm m}}\right ),
\end{equation}
where $Q=3\tilde{J}/2\tilde{I}=0.094$. In some cases, the small
initial perturbation could increase to a large value so that
non-linear effects can no longer be ignored. As postulated by Owen
et al. (1998), there may exist a saturation amplitude
$\alpha^{2}=\kappa$. When $\alpha^{2}>\kappa$, the growth of the
amplitude of the mode stops, then
\begin{equation}
 \frac{d\alpha}{dt}=0.
\end{equation}
In this stage, the spin evolution will be described as
 \begin{equation}
\frac{d\Omega}{dt}=\frac{2\Omega}{\tau_{\rm g}}\frac{\kappa
Q}{1-\kappa Q}-\frac{\Omega}{\tau_{\rm m}}\frac{1}{1-\kappa Q}.
\end{equation}
Of fundamental importance in judging the astrophysical relevance of
the instability is the determination of the saturation amplitude of
the mode. Lindblom, Tohline \& Vallisneri (2001, 2002) suggested
that the non-linear saturation amplitude may be set by dissipation
of energy in the production of shock waves. However, Gressman et al.
(2002) argued that the decay of the amplitude of the order unity is
due to leaking of energy into other fluid modes, leading to a
differential rotation configuration. Afterward, the coupling between
the {\it r}-modes and other modes was analyzed (Arras et ale 2003).
On the other hand, the role of differential rotation in the
evolution of the {\it r}-mode was also studied thoroughly (S$\rm
\acute{a}$ \& Tom$\rm \acute{e}$ 2005). All these works obtained a
credible conclusion that the saturation amplitude of the {\it
r}-mode may be not larger than the small value of $10^{-3}$. Thus,
we take $\kappa=10^{-6}$ in our calculations. The saturation
amplitude is important for the evolution of CSSs because the {\it
r-}mode in CSSs will retain the saturated state for a long time.
However, to a certain extent, the uncertainty of the saturation
value can only influence our numerical results quantitatively, but
cannot change the conclusions essentially. Note that, for NSSs, this
artificial cut of the amplitude is needless, as is also claimed by
Andersson et al. (2002).


\section{Thermal evolution}
An SS can be divided roughly into an inner, approximately
isothermal, quark core and a thin, solid, nuclear crust (Alcock,
Farhi \& Olinto 1986). For the relationship between the interior
temperature $T$ and the surface temperature $T_{s}$, we apply the
result given by Gudmundsson, Pethick \& Epstein (1983), which is
valid for a crust with the density at the base that is just larger
than $10^{8}\rm g\hspace{0.1cm}cm^{-3}$. Thus, the emissivity due to
the surface photon emission can be written as
\begin{equation}
\dot{E}_{\rm photon}=1.24{\times}10^{22}T_9^{2.2} {\rm
\hspace{0.1cm} erg \hspace{0.1cm} cm^{-2} \hspace{0.1cm} s^{-1}}.
\end{equation}
Besides the contribution from this thermal radiation, SSs also cool
via internal neutrino emission. In strange quark matter, there are
three main neutrino processes: (i) direct Urca processes
$d{\rightarrow}ue\bar{\nu}$ and $ue{\rightarrow}d{\nu}$; (ii)
modified Urca processes $dq{\rightarrow}uqe\bar{\nu}$ and
$uqe{\rightarrow}dq\nu$; (iii) the quark bremsstrahlung processes
$q_1q_2{\rightarrow}q_1q_2{\nu}\bar{\nu}$. The corresponding
neutrino emissivities read respectively (Iwamoto, 1982)
\begin{equation}
\dot{E}^{\rm(D)}_{\rm neutrino}=8.8{\times}10^{26} {\alpha}_{\rm
c}\left({\frac{{\rho}_{\rm b}}{{\rho}_0}}\right)Y_e^{1/3}T_9^6 {\rm
\hspace{0.1cm} erg \hspace{0.1cm} cm^{-3} \hspace{0.1cm} s^{-1}},
\end{equation}
\begin{equation}
\dot{E}^{\rm(M)}_{\rm neutrino}=2.83{\times}10^{19} {\alpha}_{\rm
c}^2\left({\frac{{\rho}_{\rm b}}{{\rho}_0}}\right)T_9^8{\rm
\hspace{0.1cm} erg \hspace{0.1cm} cm^{-3} \hspace{0.1cm} s^{-1}},
\end{equation}
\begin{equation}
\dot{E}^{\rm(B)}_{\rm neutrino}=2.98{\times}10^{19}
\left({\frac{{\rho}_{\rm b}}{{\rho}_0}}\right)T_9^8{\rm
\hspace{0.1cm} erg \hspace{0.1cm} cm^{-3} \hspace{0.1cm} s^{-1}},
\end{equation}
where ${\rho}_{\rm b}$ is the baryon number density and
${\rho}_0=0.17{\rm \hspace{0.1cm} fm^{-3}}$ is the nuclear
saturation density. $Y_{\rm e}={\rho}_{\rm e}/{\rho}_{\rm b}$ is the
electron fraction. On the other hand, SSs also can be heated. During
GR, in addition to the losing of angular momentum, the {\it r-}modes
also can lose energy via GR and neutrino emission (from bulk
viscosity) and can deposit energy into the thermal state of the star
via shear viscous dissipation (Owen et al. 1998). As defined in
equation (6), for the $l=m=2$ current multipole, the energy of the
{\it r-}mode is $\tilde{E}={1\over
2}\alpha^{2}\tilde{J}MR^{2}\Omega^{2}$, and the heating rate is
\begin{equation}
H_{\rm sv}={2\tilde{E}\over \tau_{\rm
sv}}=\frac{\alpha^{2}\tilde{J}MR^{2}\Omega^{2}}{\tau_{\rm sv}}
\end{equation}
Thus the thermal evolution equation is written as
\begin{equation}
C\frac{d T}{d t}=-L_{\rm neutrino}-L_{\rm photon}+H_{\rm sv},
\end{equation}
where $L_{\rm neutrino}$ and $L_{\rm photon}$ are the neutrino
luminosity and the surface photon luminosity respectively, and $C$
is the total specific heat, which is mainly contributed by quarks
and electrons (Blaschke et al. 2000):
\begin{equation}
c_{\rm q}{\simeq}2.5{\times}10^{20}\left({\frac{{\rho}_{\rm
b}}{{\rho}_0}}\right)^{2/3}T_9{\rm \hspace{0.1cm} erg \hspace{0.1cm}
cm^{-3}\hspace{0.1cm} K^{-1}},
\end{equation}
\begin{equation}
c_{\rm e}{\simeq}0.6{\times}10^{20}\left({\frac{{Y_{\rm
e}{\rho}_{\rm b}}}{{\rho}_0}}\right)^{2/3}T_9{\rm \hspace{0.1cm} erg
\hspace{0.1cm} cm^{-3}\hspace{0.1cm} K^{-1}}.
\end{equation}
In the calculation for CSSs, the reductive factor due to the pairing
gap should be added in equations (23)-(25) and (28). In other words,
the electron specific heat, the surface photon emission and the
heating term become more important for the thermal evolution of
CSSs.


\section{evolution curves}In our
calculations, we take the initial temperature $T_0=10^{10}$K, the
initial angular velocity $\Omega_0=\frac{2}{3}\sqrt{\pi G
\bar{\rho}}$, and a representative set of parameters of $Y_{\rm
e}=10^{-5}$, $\alpha_{\rm c}=0.2, m=100$MeV for SSs.\footnote{We
provide a brief review of the relevant properties of strange quark
matter in the previous sections in order to discuss the evolution
characteristic of SSs in this section. However, a more comprehensive
investigation of these properties is needed to compare the
theoretical curve with the observational data. Comments on an early
version of this paper helped us to note that the contribution to
viscosity in the CFL phase from superfluid phonons may be more
important than that of electrons (Manuel, Dobada \& Llanes-Estrada
2005), although the physical inputs left unmentioned in this paper
may not affect the general astronomical picture.}
%
\begin{figure}
\resizebox{\hsize}{!}{\includegraphics{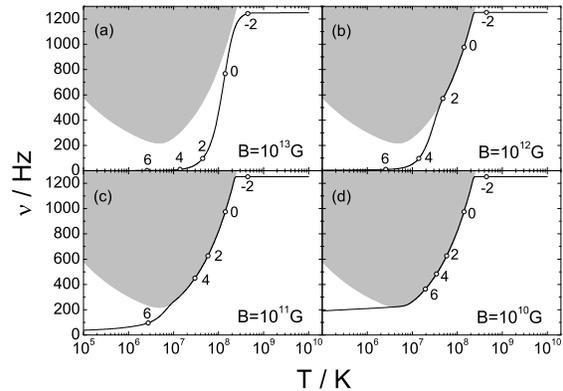}} \caption{The
evolution of NSSs (with canonical parameters $M=1.4M_{\odot}$ and
$R=10$ km) for different magnetic fields. The labels for the open
cycles represent the value of $\log(t~\rm yr^{-1})$.}
\end{figure}
%
%
%
\begin{figure}
\resizebox{\hsize}{!}{\includegraphics{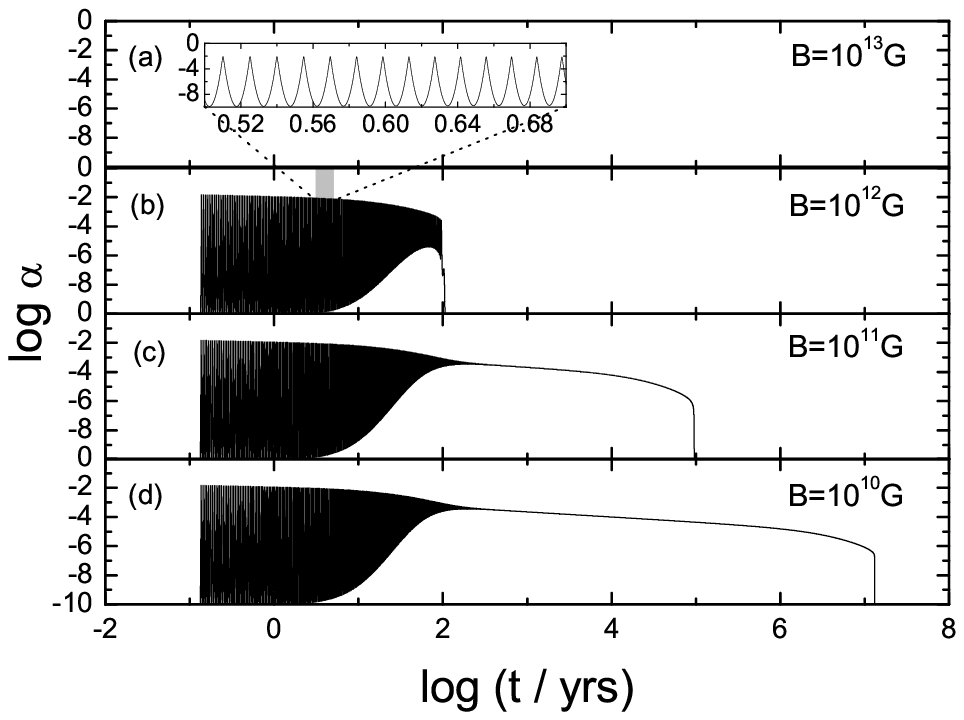}} \caption{The
evolution of the amplitude of the {\it r-}mode in NSSs for different
magnetic fields.}
\end{figure}
%
%
%
\begin{figure}
\resizebox{\hsize}{!}{\includegraphics{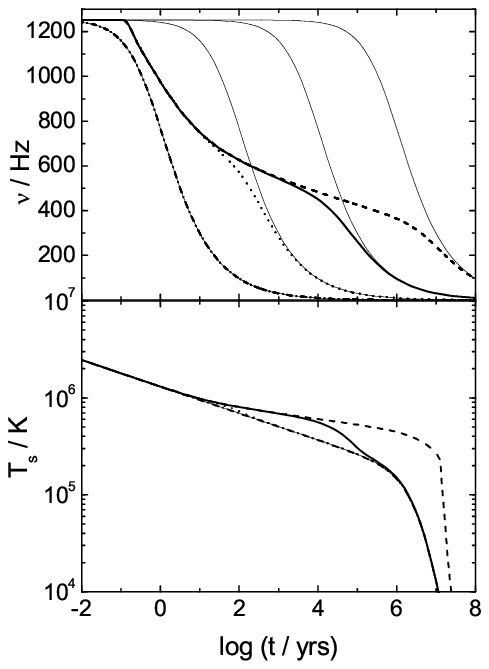}} \caption{The upper
panel shows the spin evolution of NSSs for different magnetic
fields: $B=10^{13}$G (dash-dotted curve); $B=10^{12}$G (dotted
curve); $B=10^{11}$G (solid curve); $B=10^{10}$G (dashed curve); and
the thin solid curves represent the spin-down induced by MDR purely
corresponding to the same field cases from left to right. The lower
panel shows the surface temperature evolution of NSSs for the cases
in the upper panel, and the thin solid curve corresponds to cooling
in the absence of {\it r-}mode instability.}
\end{figure}
%
%
\begin{figure}
\resizebox{\hsize}{!}{\includegraphics{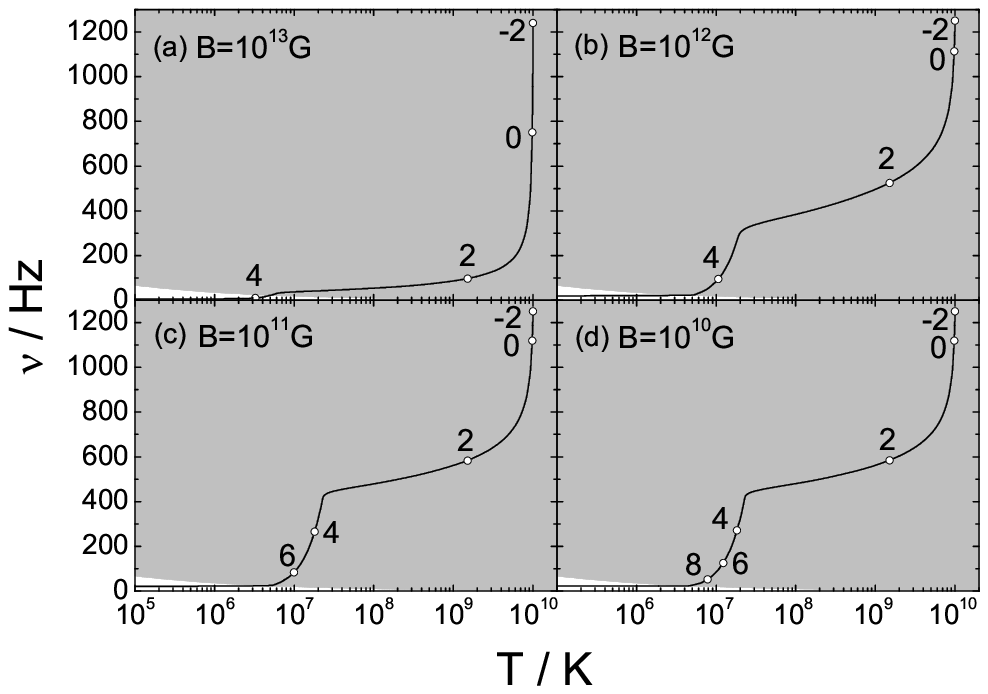}} \caption{The
evolution of CSSs (with canonical parameters $M=1.4M_{\odot}$ and
$R=10$ km) for different magnetic fields. The labels for the open
cycles represent the value of $\log(t~\rm yr^{-1})$.}
\end{figure}
%
%
%
\begin{figure}
\resizebox{\hsize}{!}{\includegraphics{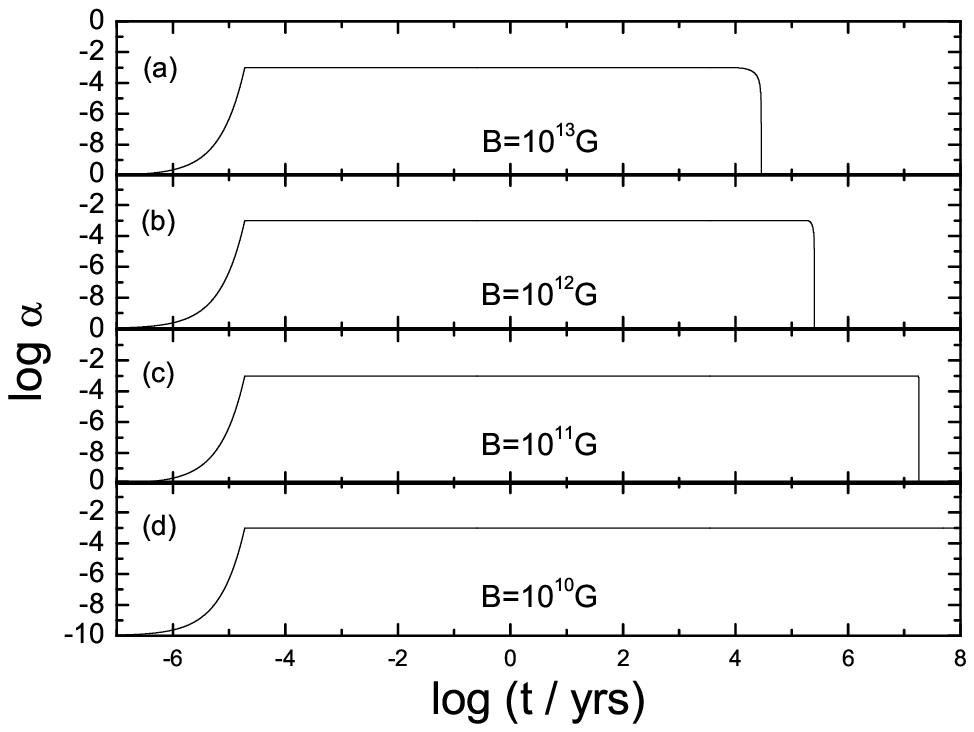}} \caption{The
evolution of the amplitude of the {\it r-}mode in CSSs for different
magnetic fields.}
\end{figure}
%
%
%
\begin{figure}
\resizebox{\hsize}{!}{\includegraphics{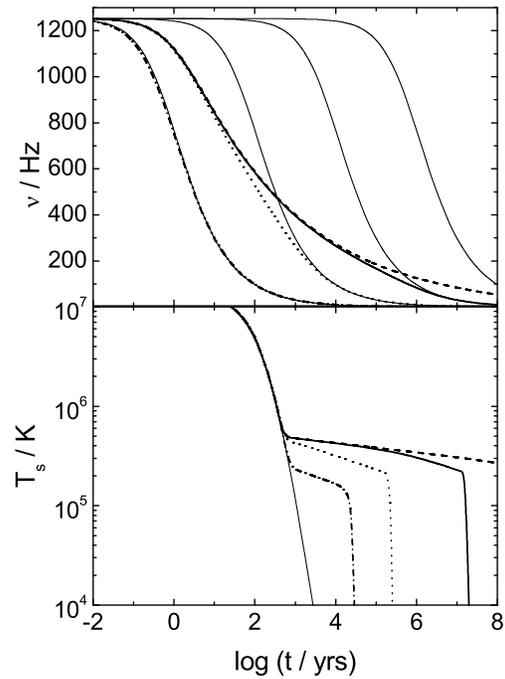}} \caption{The upper
panel shows the spin evolution of CSSs for different magnetic
fields: $B=10^{13}$G (dash-dotted curve); $B=10^{12}$G (dotted
curve); $B=10^{11}$G (solid curve); $B=10^{10}$G (dashed curve); and
the thin solid curves represent the spin-down induced by MDR purely
corresponding to the same field cases from left to right. The lower
panel shows the surface temperature evolution of CSSs for the cases
in the upper panel, and the thin solid curve corresponds to cooling
in the absence of {\it r-}mode instability.}
\end{figure}
%
\par
We firstly discuss the evolution of NSSs. The stellar evolution
tracks in the $\nu-T$ plane for different magnetic fields are shown
in Fig. 1 and several typical ages are marked on the tracks using
open cycles. Fig. 2 shows the amplitude evolution of the {\it
r-}mode with time corresponding to the same field cases. We can see
from Fig. 1(b), (c) and (d), just as calculated by Andersson et al.
(2002) for $B=0$, that the stars for non-ultrastrong ($B<10^{13}$G)
field cases will reach the instability windows several days after
birth as a result of temperature decrease and will then evolve along
the boundary of the windows. However, we also find that the specific
history of the stars is sensitively dependent on the magnetic field
intensity. It can be seen from the labelled ages that, the weaker
the field, the slower the spin-down and the cooling. Especially with
very weak ($B\leq10^{10}$G) fields, the stars can remain at a very
high spin frequency, such as 400Hz, even at the age of over a
million years. Finally, the evolution tracks will depart from the
instability windows at the moment that the magnetic braking effect
exceeds that of GR (see the upper panel of Fig.3). Thus, the
stronger the field, the earlier the departure. Synchronously, this
departure also leads to the end (a catastrophically decay) of the
{\it r-}mode as shown in Fig. 2. Before this abrupt drop, however,
the variation of the mode is small apart from the first several
decades, during which the amplitude of the {\it r-}mode oscillates
greatly until it converges to a value of $\sim10^{-3}$ as discussed
by Andersson et al. (2002). At the extreme, a star with an
ultrastrong ($B>10^{13}$G) field decelerates so rapidly due to MDR
that the track will not come into contact with the instability
window as shown in Fig. 1(a). Thus, no {\it r-}mode arises in the
star throughout its life as shown in Fig. 2(a). To be clear, Fig. 3
shows the evolution curves of spin frequency (upper panel) and
surface temperature (lower panel) corresponding to the cases shown
in Fig. 1. Not surprising, being governed by Eq. (19), the spin
history of the stars with non-ultrastrong ($B<10^{13}$G) fields can
be divided into two stages: (i) the stage due to angular momentum
transfer by viscosity (losing angular momentum via GR ultimately) at
early ages in the presence of the {\it r-}mode; and (ii) the stage
due to MDR in the absence of the {\it r-}mode. On the other hand, as
shown in the lower panel, the cooling of the stars with weak fields
is delayed under the impact of heating due to the viscous
dissipation of the {\it r-}mode, whereas the curve of the strong
field superposes the curve without the consideration of the {\it
r-}mode instability.
\par
In the second situation, colour superconductivity suppresses the
cooling effect extremely, whereas the heating effect is enhanced
considerably. The evolution tracks of CSSs in the $\nu-T$ plane for
different magnetic fields and the corresponding amplitude evolution
curves of the {\it r-}mode are shown in Fig. 4 and 5, respectively.
The same as Figs 3 but for CSSs are also shown in Fig. 6. Differing
from NSSs, the {\it r-}mode in CSSs can appear at the birth of the
stars and quickly ($\sim1000$s, as shown clearly in Fig. 5) rise to
the saturated state because the bulk viscosity is blocked. However,
at the same time, the temperature nearly does not decrease because
the neutrino cooling effect is suppressed. However, because the
specific heat is only contributed from electrons, the temperature
decreases finally in the following several hundred years until the
photon luminosity is nearly compensated by the heat gain from the
shear viscous dissipation of the {\it r-}mode. From then on, the
stars should experience a long-term slow cooling stage. As discussed
by Blaschke et al. (2000) and Yu \& Zheng (2006) previously, CSSs
will become very cold at early ages ($<1000$yrs) if the heating
effect is ignored. However, the presence of the long-term slow
cooling stage can change the situation completely as shown in Fig. 6
(lower panel). We can see the surface temperature can remain at a
high value of $\sim10^{6}$K lasting several million years. The
spin-down of CSSs that is shown in the upper panel of Fig. 6 is
similar to the one of NSSs, but the stage dominated by GR is much
longer because the life of the {\it r-}mode in CSSs is longer as
shown in Fig.5.


\section{Conclusion and discussions}
We have studied the evolution of SSs synthetically, considering the
influence of {\it r-}mode instability. The spin history is divided
into two stages and the heating effect due to viscous dissipation of
the {\it r-}mode can delay the cooling of SSs with non-ultrastrong
fields remarkably. Our prime objective is to investigate the
behaviour of SSs in colour superconducting phase (to be specific,
the CFL phase). This problem is estimated simply by Madsen (2000),
who argued that CSSs should spin-down to a very low spin frequency
for hours due to the {\it r-}mode instability. Conversely, our
results indicate that the stars can be located at the bottom of the
window for millions of years due to their characteristic cooling
behaviour. A CSS could be a rapid revolving pulsar. Differing from
NSSs, the cooling of CSSs is changed completely by heating due to
{\it r-}mode damping. Except for the first several hundred years,
CSSs cool very slowly. In other words, the stars can remain at a
high temperature over several million years but do not become very
cold at early ages as calculated in the absence of the heating
effect. Thus, we believe that it is possible to detect a hot source
being a CSS but not, as argued by Blaschke et al. (2000), that CSSs
are rejected by the X-ray data.
\par
In our opinion, when we talk about stellar evolution, we must
consider the cooling and the spin-down synthetically because the
rotation of the star is a large possible energy store for the
heating of the star. We note that Reisenegger (1995) proposed a
chemical heating mechanism due to the derivative from the
$\beta-$equilibrium of neutron matter with the spin-down of neutron
stars. The same issue was discussed in NSSs by Cheng \& Dai (1996),
who presented that chemical heating delays the later ($>10^6$yrs)
cooling of NSSs significantly. In this paper, however, it is the
{\it r-}mode instability that couples the cooling and the spin-down:
the kinetic energy transforms into the canonical energy of the {\it
r-}modes via GR, then is deposited into a thermal state via the
dissipation of the {\it r-}modes. Our results show that the heating
due to the {\it r-}mode damping delays the earlier ($<10^6$yrs)
cooling remarkably. To summarize, we argue that the heating due to
spin-down plays an important role in the evolution of an NSS
throughout its life. On the other hand, for a CSS, the chemical
heating is blocked because the reactions between the quarks are
suppressed by the pairing gap. Thus, the heating due to the {\it
r-}mode damping becomes more decisive for its evolution. As
discussed above, the spin-down can impact the cooling of an SS as an
energy source. Besides this direct influence, the spin evolution
also can determine the change of other properties of the star, such
as the variation of the density of the star. We therefore think that
the spin-down can also impact the cooling behaviour indirectly.
Looking at more detailed calculations, previous modelling of thermal
evolution of SSs based on the assumption that the spin-down is only
determined by MDR needs to be reconsidered because it is GR but not
MDR that dominates the first spin-down of the star in the presence
of the {\it r-}modes.
\par
Using NSSs, several authors (Madsen, 2000; Andersson et al. 2002)
try to explain the clustering of low mass X-ray binaries in the
$\nu-T$ plane. Our results (Fig. 1) show that, for weak-field
($B\leq10^{10}$G) cases, MDR has no influence on the spin-down of
the stars that are younger than 10 million years old. This implies
the existence of a weak magnetic field in millisecond pulsars if we
follow the same philosophy that has been used by Madsen (2000).
However, in our opinion, the final clarification of this problem
should dependent on careful consideration of the accretion of stars
as Watts \& Andersson (2002) did for neutron stars.
\par
\textit{\textbf{Acknowledgements}}. We would like to thank for the
support by the National Natural Science Foundation of China, grant
nos 10373007 and 90303007, and the Ministry of Education of China,
with project nos 704035, 20040511005. We also thank Cristina Manuel
for significant comments on the viscosity.

\label{lastpage}

\begin{thebibliography}{}

\bibitem[\protect\citeauthoryear{{Abraham}, {Crawford} \& {McHardy}}
  {{Abraham}, {Crawford} \& {McHardy}}{1992}]
  {abraham+92}
  {Abraham} R.~G.,  {Crawford} C.~S.,    {McHardy} I.~M.,  1992, 
  \apj, 401, 474


\bibitem[\protect\citeauthoryear{{Bahcall}, {Kirhakos}, {Saxe} \& {Schneider}}
  {{Bahcall} et~al.}{1997}]
  {bahcall+97}
  {Bahcall} J.~N.,  {Kirhakos} S.,  {Saxe} D.~H.,    {Schneider} D.~P.,  1997,
  \apj, 479, 642

\bibitem[\protect\citeauthoryear{{Bernardi}, et~al.}
  {{Bernardi} et~al.}{2003}]
  {bernardi+03}
  {Bernardi} M., et~al., 
  2003, \aj, 125, 1849
  
\bibitem[\protect\citeauthoryear{{Block} \& {Stockton}}
  {{Block} \& {Stockton}}{1991}]
  {blockstockton91}
  {Block} D.~L.,  {Stockton} A.,  1991, 
  \aj, 102, 1928

\bibitem[\protect\citeauthoryear{{Blundell}, {Beasley}, {Lacy} \& {Garrington}}
  {{Blundell} et~al.}{1996}]
  {blundell+96}
  {Blundell} K.~M.,  {Beasley} A.~J.,  {Lacy} M.,    {Garrington} S.~T.,  1996,
  \apjl, 468, L91

\bibitem[\protect\citeauthoryear{{Blundell} \& {Rawlings}}{{Blundell} \&
    {Rawlings}}{2001}]{blundellrawlings01}
  {Blundell} K.~M.,  {Rawlings} S.,  2001, 
  \apjl, 562, L5

\bibitem[\protect\citeauthoryear{{Boyce}, et~al.}
  {{Boyce} et~al.}{1998}]
  {boyce+98}
  {Boyce} P.~J., et~al.,
  1998, \mnras, 298, 121

\bibitem[\protect\citeauthoryear{{Boyce}, {Disney} \& {Bleaken}}{{Boyce}
    et~al.}{1999}]{boyce+99}
  {Boyce} P.~J.,  {Disney} M.~J.,    {Bleaken} D.~G.,  1999, \mnras, 302, L39

\bibitem[\protect\citeauthoryear{{Dunlop}, {McLure}, {Kukula}, {Baum}, {O'Dea}
  \& {Hughes}}{{Dunlop} et~al.}{2003}]{dunlop+03}
{Dunlop} J.~S.,  {McLure} R.~J.,  {Kukula} M.~J.,  {Baum} S.~A.,  {O'Dea}
  C.~P.,    {Hughes} D.~H.,  2003, \mnras, 340, 1095

\bibitem[\protect\citeauthoryear{{Efstathiou}, {Ellis} \&
  {Peterson}}{{Efstathiou} et~al.}{1988}]{EEP88}
{Efstathiou} G.,  {Ellis} R.~S.,    {Peterson} B.~A.,  1988, \mnras, 232, 431

\bibitem[\protect\citeauthoryear{{Falomo}, {Kotilainen} \& {Treves}}{{Falomo}
  et~al.}{2001}]{falomo+01}
{Falomo} R.,  {Kotilainen} J.,    {Treves} A.,  2001, \apj, 547, 124

\bibitem[\protect\citeauthoryear{{Ferrarese} \& {Merritt}}{{Ferrarese} \&
  {Merritt}}{2000}]{ferraresemerritt00}
{Ferrarese} L.,  {Merritt} D.,  2000, \apjl, 539, L9

\bibitem[\protect\citeauthoryear{{Gebhardt}, et~al.}
  {{Gebhardt} et~al.}{2000}]
  {gebhardt+00}
  {Gebhardt} K.,  et~al.,
  2000, \apjl, 539, L13

\bibitem[\protect\citeauthoryear{{Goldschmidt}, {Kukula}, {Miller} \&
  {Dunlop}}{{Goldschmidt} et~al.}{1999}]{goldschmidt+99}
{Goldschmidt} P.,  {Kukula} M.~J.,  {Miller} L.,    {Dunlop} J.~S.,  1999,
  \apj, 511, 612

\bibitem[\protect\citeauthoryear{{Goldschmidt}, {Miller}, {La Franca} \&
  {Cristiani}}{{Goldschmidt} et~al.}{1992}]{goldschmidt+92}
{Goldschmidt} P.,  {Miller} L.,  {La Franca} F.,    {Cristiani} S.,  1992,
  \mnras, 256, 65P

\bibitem[\protect\citeauthoryear{{Green} \& {Yee}}{{Green} \&
  {Yee}}{1984}]{greenyee84}
{Green} R.~F.,  {Yee} H.~K.~C.,  1984, \apjs, 54, 495

\bibitem[\protect\citeauthoryear{{Gregory}, {Vavasour}, {Scott} \&
  {Condon}}{{Gregory} et~al.}{1994}]{gregory+94}
{Gregory} P.~C.,  {Vavasour} J.~D.,  {Scott} W.~K.,    {Condon} J.~J.,  1994,
  \apjs, 90, 173

\bibitem[\protect\citeauthoryear{{Hamilton}, {Casertano} \&
  {Turnshek}}{{Hamilton} et~al.}{2002}]{hamilton+02}
{Hamilton} T.~S.,  {Casertano} S.,    {Turnshek} D.~A.,  2002, \apj, 576, 61

\bibitem[\protect\citeauthoryear{{Hooper}, {Impey} \& {Foltz}}{{Hooper}
  et~al.}{1997}]{hooper+97}
{Hooper} E.~J.,  {Impey} C.~D.,    {Foltz} C.~B.,  1997, \apjl, 480, L95

\bibitem[\protect\citeauthoryear{{Hutchings}}{{Hutchings}}{1987}]{hutchings87}
{Hutchings} J.~B.,  1987, \apj, 320, 122

\bibitem[\protect\citeauthoryear{{Hutchings}, {Frenette}, {Hanisch}, {Mo},
  {Dumont}, {Redding} \& {Neff}}{{Hutchings} et~al.}{2002}]{hutchings+02}
{Hutchings} J.~B.,  {Frenette} D.,  {Hanisch} R.,  {Mo} J.,  {Dumont} P.~J.,
  {Redding} D.~C.,    {Neff} S.~G.,  2002, \aj, 123, 2936

\bibitem[\protect\citeauthoryear{{Hutchings}, {Johnson} \& {Pyke}}{{Hutchings}
  et~al.}{1988}]{hutchings+88}
{Hutchings} J.~B.,  {Johnson} I.,    {Pyke} R.,  1988, \apjs, 66, 361

\bibitem[\protect\citeauthoryear{{Hutchings} \& {Neff}}{{Hutchings} \&
  {Neff}}{1990}]{hutchings+90}
{Hutchings} J.~B.,  {Neff} S.~G.,  1990, \aj, 99, 1715

\bibitem[\protect\citeauthoryear{{Hutchings} \& {Neff}}{{Hutchings} \&
  {Neff}}{1991}]{hutchingsneff91}
{Hutchings} J.~B.,  {Neff} S.~G.,  1991, \aj, 101, 2001

\bibitem[\protect\citeauthoryear{{Hutchings} \& {Neff}}{{Hutchings} \&
  {Neff}}{1992}]{hutchingsneff92}
{Hutchings} J.~B.,  {Neff} S.~G.,  1992, \aj, 104, 1

\bibitem[\protect\citeauthoryear{{J{\o}rgensen}, {Franx} \&
  {Kjaergaard}}{{J{\o}rgensen} et~al.}{1996}]{jorgensen+96}
{J{\o}rgensen} I.,  {Franx} M.,    {Kjaergaard} P.,  1996, \mnras, 280, 167

\bibitem[\protect\citeauthoryear{{Kormendy} \& {Gebhardt}}{{Kormendy} \&
  {Gebhardt}}{2001}]{kormgeb01}
{Kormendy} J.,  {Gebhardt} K.,  2001, in 20th Texas Symposium on relativistic
  astrophysics {Supermassive Black Holes in Galactic Nuclei (Plenary Talk)}.

\bibitem[\protect\citeauthoryear{{Krist}}{{Krist}}{1999}]{tinytim}
{Krist} J.,  1999, TinyTim User Manual

\bibitem[\protect\citeauthoryear{{Kukula}, {Dunlop}, {McLure}, {Miller},
  {Percival}, {Baum} \& {O'Dea}}{{Kukula} et~al.}{2001}]{kukula+01}
{Kukula} M.~J.,  {Dunlop} J.~S.,  {McLure} R.~J.,  {Miller} L.,  {Percival}
  W.~J.,  {Baum} S.~A.,    {O'Dea} C.~P.,  2001, \mnras, 326, 1533

\bibitem[\protect\citeauthoryear{{Lehnert}, {van Breugel}, {Heckman} \&
  {Miley}}{{Lehnert} et~al.}{1999}]{lehnert+99b}
{Lehnert} M.~D.,  {van Breugel} W.~J.~M.,  {Heckman} T.~M.,    {Miley} G.~K.,
  1999, \apjs, 124, 11

\bibitem[\protect\citeauthoryear{{M{\' a}rquez}, {Petitjean}, {Th{\' e}odore},
  {Bremer}, {Monnet} \& {Beuzit}}{{M{\' a}rquez} et~al.}{2001}]{marquez+01}
{M{\' a}rquez} I.,  {Petitjean} P.,  {Th{\' e}odore} B.,  {Bremer} M.,
  {Monnet} G.,    {Beuzit} J.-L.,  2001, \aap, 371, 97

\bibitem[\protect\citeauthoryear{{Magorrian}, et~al.}
  {{Magorrian} et~al.}{1998}]
  {magorrian+98}
  {Magorrian} J., et~al., 
  1998, \aj, 115, 2285

\bibitem[\protect\citeauthoryear{{Malkan}}{{Malkan}}{1984}]{malkan84}
{Malkan} M.~A.,  1984, \apj, 287, 555

\bibitem[\protect\citeauthoryear{{Marconi}, {Axon}, {Macchetto}, {Capetti},
  {Soarks} \& {Crane}}{{Marconi} et~al.}{1997}]{marconi+97}
{Marconi} A.,  {Axon} D.~J.,  {Macchetto} F.~D.,  {Capetti} A.,  {Soarks}
  W.~B.,    {Crane} P.,  1997, \mnras, 289, L21

\bibitem[\protect\citeauthoryear{{Marconi} \& {Hunt}}{{Marconi} \&
  {Hunt}}{2003}]{marconihunt03}
{Marconi} A.,  {Hunt} L.~K.,  2003, \apjl, 589, L21

\bibitem[\protect\citeauthoryear{{McLeod} \& {Rieke}}{{McLeod} \&
    {Rieke}}{1995}]{mcleodrieke95} 
{McLeod} B.~A.,  {Rieke} G.~H.,  1995, \apj, 441, 96

\bibitem[\protect\citeauthoryear{{McLeod} \& {McLeod}}{{McLeod} \&
  {McLeod}}{2001}]{mcleod01}
{McLeod} K.~K.,  {McLeod} B.~A.,  2001, \apj, 546, 782

\bibitem[\protect\citeauthoryear{{McLeod}, {Rieke} \&
  {Storrie-Lombardi}}{{McLeod} et~al.}{1999}]{mcleod+99}
{McLeod} K.~K.,  {Rieke} G.~H.,    {Storrie-Lombardi} L.~J.,  1999, \apjl, 511,
  L67

\bibitem[\protect\citeauthoryear{{McLure} \& {Dunlop}}{{McLure} \&
  {Dunlop}}{2002}]{mcluredunlop02}
{McLure} R.~J.,  {Dunlop} J.~S.,  2002, \mnras, 331, 795

\bibitem[\protect\citeauthoryear{{McLure}, {Dunlop} \& {Kukula}}
  {{McLure} et~al.}{2000}]
  {mclure+00}
  {McLure} R.~J.,  {Dunlop} J.~S.,    {Kukula} M.~J.,  2000, \mnras, 318, 693

\bibitem[\protect\citeauthoryear{{McLure}, {Kukula}, {Dunlop}, {Baum}, {O'Dea}
  \& {Hughes}}{{McLure} et~al.}{1999}]{mclure+99}
{McLure} R.~J.,  {Kukula} M.~J.,  {Dunlop} J.~S.,  {Baum} S.~A.,  {O'Dea}
  C.~P.,    {Hughes} D.~H.,  1999, \mnras, 308, 377

\bibitem[\protect\citeauthoryear{{Percival}, {Miller}, {McLure} \&
  {Dunlop}}{{Percival} et~al.}{2001}]{percival+01}
{Percival} W.~J.,  {Miller} L.,  {McLure} R.~J.,    {Dunlop} J.~S.,  2001,
  \mnras, 322, 843

\bibitem[\protect\citeauthoryear{{Press}, {Teukolsky}, {Vetterling} \&
  {Flannery}}{{Press} et~al.}{1992}]{numrec}
{Press} W.~H.,  {Teukolsky} S.~A.,  {Vetterling} W.~T.,    {Flannery} B.~P.,
  1992, Numerical recipes in FORTRAN. The art of scientific computing.
Cambridge: University Press, 1992, 2nd ed.

\bibitem[\protect\citeauthoryear{{Puchnarewicz}, et~al.}
  {{Puchnarewicz} et~al.}{1992}]
  {puchnarewicz+92}
  {Puchnarewicz} E.~M.,  et~al., 
  1992, \mnras, 256, 589

\bibitem[\protect\citeauthoryear{{Reimers}, et~al.}
  {{Reimers} et~al.}{1995}]
  {reimers+95}
  {Reimers} D.,  et~al.,
  1995, \aap, 303, 449

\bibitem[\protect\citeauthoryear{{Ridgway}, {Heckman}, {Calzetti} \&
  {Lehnert}}{{Ridgway} et~al.}{2001}]{ridgway+01}
{Ridgway} S.~E.,  {Heckman} T.~M.,  {Calzetti} D.,    {Lehnert} M.,  2001,
  \apj, 550, 122

\bibitem[\protect\citeauthoryear{{S\'{e}rsic}}{{S\'{e}rsic}}{1968}]{sersic68}
{S\'{e}rsic} J.~L.,  1968, in Atlas de Galaxes Australes; Vol. Book; Page 1 {Atlas de Galaxes Australes}.

\bibitem[\protect\citeauthoryear{{Smith}, {Heckman}, {Bothun}, {Romanishin} \&
  {Balick}}{{Smith} et~al.}{1986}]{smith+86}
{Smith} E.~P.,  {Heckman} T.~M.,  {Bothun} G.~D.,  {Romanishin} W.,    {Balick}
  B.,  1986, \apj, 306, 64

\bibitem[\protect\citeauthoryear{{Stockton} \& {Ridgway}}{{Stockton} \&
  {Ridgway}}{2001}]{stocktonridgway01}
{Stockton} A.,  {Ridgway} S.~E.,  2001, \apj, 554, 1012

\bibitem[\protect\citeauthoryear{{Tadhunter}, {Marconi}, {Axon}, K., {Robinson}
  \& {Jackson}}{{Tadhunter} et~al.}{2003}]{tadhunter+03}
{Tadhunter} C.,  {Marconi} A.,  {Axon} D.,  K. W.,  {Robinson} T.~G.,
  {Jackson} N.,  2003, \mnras

\bibitem[\protect\citeauthoryear{{Veron-Cetty} \& {Woltjer}}{{Veron-Cetty} \&
  {Woltjer}}{1990}]{veron90}
{Veron-Cetty} M.~.,  {Woltjer} L.,  1990, \aap, 236, 69

\bibitem[\protect\citeauthoryear{{Veron-Cetty} \& {Veron}}{{Veron-Cetty} \&
  {Veron}}{1993}]{VCV1993}
{Veron-Cetty} M.-P.,  {Veron} P.,  1993, {A Catalogue of quasars and active
  nuclei}.
ESO Scientific Report, Garching: European Southern Observatory (ESO), |c1993,
  6th ed.

\bibitem[\protect\citeauthoryear{{Veron-Cetty} \& {Veron}}{{Veron-Cetty} \&
  {Veron}}{2000}]{VCV2000}
{Veron-Cetty} M.-P.,  {Veron} P.,  2000, {A catalogue of quasars and active
  nuclei}.
A catalogue of quasars and active nuclei / M.-P.~Veron-Cetty and P.~Veron.~
  Garching bei Munchen, Germany : European Southern Observatory,
  c2000.~(Scientific report (European Southern Observatory) ; no.~19)

\bibitem[\protect\citeauthoryear{{Voges}, {Aschenbach}, {Boller}, {Br{\"
  a}uninger}, {Briel} \& {Burkert}}{{Voges} et~al.}{1999}]{voges+99}
{Voges} W.,  {Aschenbach} B.,  {Boller} T.,  {Br{\" a}uninger} H.,  {Briel} U.,
     {Burkert} W.,  1999, \aap, 349, 389

\bibitem[\protect\citeauthoryear{{Wright}, {McHardy} \& {Abraham}}{{Wright}
  et~al.}{1998}]{wright+98}
{Wright} S.~C.,  {McHardy} I.~M.,    {Abraham} R.~G.,  1998, \mnras, 295, 799

\bibitem[\protect\citeauthoryear{{Wyckoff}, {Gehren} \& {Wehinger}}{{Wyckoff}
  et~al.}{1981}]{wyckoff+81}
{Wyckoff} S.,  {Gehren} T.,    {Wehinger} P.~A.,  1981, \apj, 247, 750

\bibitem[\protect\citeauthoryear{{Yee} \& {Green}}{{Yee} \&
  {Green}}{1987}]{yeegreen87}
{Yee} H.~K.~C.,  {Green} R.~F.,  1987, \apj, 319, 28

\end{thebibliography}


\begin{thebibliography}
\scriptsize
\bibitem{Alcock86}Alcock C., Farhi E., Olinto A., 1986, ApJ, 310,
261
\bibitem[Alford(2004)]{Alford(2004)}Alford M., 2004, J. Phys. G, 30, 441
\bibitem{Andersson98}Andersson N., ApJ, 1998, 502, 708
\bibitem{Andersson02}Andersson N., Jones D. I., Kokkotas K. D.,
2002, MNRAS, 337, 1224
\bibitem{Arras03}Arras P., Flanagan E. E., Morsink S. M., Schenk A. K.,
Teukolsky S.A., Wasserman I., 2003, ApJ, 591, 1129
\bibitem{Blaschke00}Blaschke D., Kl\"{a}hn T., Voskresensky D. N., 2000, ApJ, 533,
406
\bibitem{Cheng1996}Cheng K. S., Dai Z. G., 1996, ApJ, 468, 819
\bibitem{Frindman98}Frindman J. L., Morsink S. M., 1998, ApJ, 502,
714
\bibitem{Friedman1978}Friedman J. L., Schutz B. F., 1978, ApJ, 222,
281
\bibitem{Gressman02}Gressman P., Lin L. M., Suen W. M., Stergioulas N., Friedman J.L., 2002, Phys. Rev. D, 66,
041303
\bibitem{Gudmundsson83}Gudmundsson E. H., Pethick C. J., Epstein R. I., 1983, ApJ, 272,
286
\bibitem{Gusakov05}Gusakov M. E., Yakovlev D. G., Gnedin O. Y.,
2005, MNRAS, 361, 1415
\bibitem{Ho2000}Ho W. C. G., Lai D., 2000, ApJ, 543, 386
\bibitem{Iwamoto82}Iwamoto N., 1982, Ann. Phys., 141, 1
\bibitem{Lindblom98}Lindblom L., Owen B. J., Morsink S. M., 1998,
Phys. Rev. Lett., 80, 4843
\bibitem{Lindblom99}Lindblom L., Mendell G., Owen B. J., 1999,
Phys. Rev. D, 60, 064006
\bibitem{Lindblom01}Lindblom L., Tohline J. E., Vallisneri M.,
2001, Phys. Rev. Lett., 86, 1152
\bibitem{Lindblom02}Lindblom L., Tohline J. E., Vallisneri M.,
2002, Phys. Rev. D, 65, 084039
\bibitem{Madsen2000}Madsen J., 2000, Phys. Rev. Lett., 85, 10
\bibitem{Manuel2005}Manuel C., Dobado A., Llanes-Estrada F. J., 2005, JHEP, 0509, 076 [arXiv: hep-ph/0406058]
\bibitem{Owen98}Owen B. J., Lindblom L., Cutler C., Schutz B. F.,
Vecchio A., Andersson N., 1998, Phys. Rev. D, 58, 084020
\bibitem{Papaloizou1978}Papaloizou J., Pringle J. E., 1978, MNRAS,
182, 423
\bibitem{Provost81}Provost J., Berthomieu G., Rocca A., 1981,
A\&A, 94, 126
\bibitem{Reisenegger1995}Reisenegger A., 1995, ApJ, 442, 749
\bibitem{Sa}S$\rm \acute{a}$ P. M., Tom$\rm \acute{e}$ B., 2005, Phys. Rev. D, 71,
044007
\bibitem[Shovkovy(2004)]{Shovkovy(2004)}Shovkovy I. A., 2005, Found. Phys., 35, 1309
\bibitem{Watts2002}Watts A. L., Andersson N., 2002, MNRAS, 333,
943
\bibitem{Yu2004}Yu Y. W., Zheng X. P., 2006, A\&A, 450, 1071

\end{thebibliography}
\end{document}